\documentclass[aps,showpacs,twocolumn,superscriptaddress]{revtex4}

\usepackage{graphicx}
\usepackage{epsfig}
\usepackage{amsfonts}
\usepackage{amsmath,amssymb}

\newcommand{\mc}{\hspace{0.84em}}

\begin{document}

\title{Spin-orbit and tensor mean-field effects on spin-orbit
splitting including self-consistent core polarizations}

\author{M. Zalewski}
\affiliation{Institute of Theoretical Physics, University of Warsaw, ul. Ho\.za
69, 00-681 Warsaw, Poland.}

\author{J. Dobaczewski}
\affiliation{Institute of Theoretical Physics, University of Warsaw, ul. Ho\.za
69, 00-681 Warsaw, Poland.}
\affiliation{Department of Physics, P.O. Box 35 (YFL),
FI-40014 University of Jyv\"askyl\"a, Finland}

\author{W. Satu{\l}a}
\affiliation{Institute of Theoretical Physics, University of Warsaw, ul. Ho\.za
69, 00-681 Warsaw, Poland.}

\author{T.R. Werner}
\affiliation{Institute of Theoretical Physics, University of Warsaw, ul. Ho\.za
69, 00-681 Warsaw, Poland.}

\date{\today}

\begin{abstract}
A new strategy of fitting the coupling constants of the nuclear
energy density functional is proposed, which shifts attention
from ground-state bulk to single-particle properties. The latter
are analyzed in terms of the bare single-particle energies and
mass, shape, and spin core-polarization effects.
Fit of the isoscalar spin-orbit and both
isoscalar and isovector tensor coupling constants directly to
the $f_{5/2}-f_{7/2}$ spin-orbit splittings in $^{40}$Ca, $^{56}$Ni,
and $^{48}$Ca is proposed as a practical realization of this new
programme. It is shown that this fit requires drastic
changes in the isoscalar spin-orbit strength and the tensor coupling
constants as compared to the commonly accepted values but it
considerably and systematically improves basic single-particle
properties including
spin-orbit splittings and magic-gap energies.
Impact of these changes on nuclear binding energies is also
discussed.

\end{abstract}

\pacs{21.10.Hw, 21.10.Pc, 21.60.Cs, 21.60.Jz} \maketitle

\section{Introduction}
\label{sec1}

In this work we propose and explore two new ideas pertaining to the
energy density functional (EDF) methods.
First, we suggest a necessity of shifting attention and focus
of these methods from ground-state bulk properties (e.g.\ total
nuclear masses) to single-particle (s.p.) properties, and to look for
a spectroscopic-quality EDFs that would correctly describe nuclear
shell structure. Proper positions of s.p.\ levels are instrumental
for good description of deformation, pairing, particle-core coupling,
and rotational effects, and many other phenomena.

On the one hand, careful adjustment of these positions were at the
heart of tremendous success of phenomenological mean-field (MF) models,
like those of Nilsson, Woods-Saxon, or folded Yukawa \cite{[RS80]}.
On the other hand, similarly successful phenomenological description
of nuclear masses, within the so-called microscopic-macroscopic
method \cite{[Lun03]}, relies on the liquid-drop mass surface, which
is entirely decoupled from the s.p.\ structure.

Up two now, methods based on using EDFs, in any of its variants like
local Skyrme, non-local Gogny, or relativistic-mean-field (RMF)
\cite{[Ben03]} approach, were mostly using adjustments to bulk
nuclear properties. As a result, shell properties were described
poorly. After so many years of investigations, a further increase in
precision and predictability of all methods based on the EDFs may
require extensions beyond forms currently in use. Before this can be
fully achieved, we propose to first take care of the s.p.\
properties, and come back to precise adjustment of bulk properties
once these extensions are implemented.

Second, we propose to look at the s.p.\ properties of nuclei through
the magnifying glass of odd-even mass differences. This idea has
already been put forward in a seminal paper by Rutz {\em et al.}
\cite{[Rut99a]}, where calculations performed within the
RMF approach were presented. In this paper we
perform analogous analysis for the EDFs based on the Skyrme
interactions.

On the one hand,
it was recognized long time ago, cf., e.g.,
Refs.~\cite{[Ham76aw],[Ber80w]}, that the theoretical s.p.\ energies,
defined as eigenvalues of the MF Hamiltonian, cannot be
directly compared to experiment, because they are strongly
renormalized by the particle-core coupling. On the other hand,
procedures used to deduce the s.p.\ energies from experiment
\cite{[Isa02w],[Oro96w],[Sch07a]} require various theoretical
assumptions, by which these quantities cease to result from direct
experimental observation. In the past, these theoretical and
experimental caveats hampered the use of s.p.\ energies for proper
adjustments of EDFs. However, the odd-even mass differences carry
very similar physical information to that given by s.p.\ energies,
and have advantage of being clearly defined, both experimentally and
theoretically.

Indeed, experimental difference in mass between an odd nucleus and
its lighter even-even neighbor, i.e., the particle separation energy,
is an easily available and unambiguous piece of data, which reflects
the physical role of the s.p.\ energy, with all polarizations and
couplings taken into account. Similarly, differences of masses
between the low-lying excited states in an odd system and lighter
even-even neighbor may illustrate effective positions of higher s.p.\
states. Physical connections between these mass differences and s.p.\
energies are closest in semi-magic nuclei, which will be studied in the
present paper.

In theory, the primary goal of the EDF methods is to describe
ground-state energies of fermion systems, i.e., in nuclear-physics
applications -- masses of nuclei. For odd systems, the EDF methods
should give masses of ground states and of several low-lying excited
states; the latter being obtained by blocking specific s.p.\
orbitals. We stress here that in an odd system, separate
self-consistent calculations have to be performed for each of the
blocked states, so as to allow the system for exploring all possible
polarizations exerted by the odd particle on the even-even core.
Again, this procedure is clearly defined and entirely within the
scope and spirit of the EDF method, which is supposed to provide
exact energies of correlated states. Note, that although one here
calculates {\em total masses} of odd and even systems, the comparison
with experiment involves the {\em differences of masses}, for which
many effects cancel out. Therefore, one can confidently attribute
calculated differences of masses to properties of effective s.p.\
energies, with all polarization effects taken into account, like in
the experiment. Again, in semi-magic nuclei such a connection is most
reliable.

Following the above line of reasoning, in the present paper we study
experimental and theoretical aspects of the spin-orbit (SO) splitting
between the s.p.\ states. In particular, we analyze the role of the SO
and tensor MFs in providing the correct values of the SO
splittings across the nuclear chart. The paper is organized as
follows: In Sect.~\ref{sec2} we briefly recall basic theoretical building
blocks related to the SO and tensor terms of the EDFs and interactions.
In Sect.~\ref{sec3} we discuss in details three major sources of the
core polarization: the mass, shape, and spin polarizabilities.
In Sect.~\ref{sec4} we present a novel method that
allows for a firm adjustment of the SO and tensor
coupling constants arguing that
currently used functionals require major revisions
concerning strengths of both these terms.
The analysis is based on the $f_{7/2} - f_{5/2}$ SO splittings in
three key nuclei, including
spin-saturated isoscalar $^{40}$Ca, spin-unsaturated isoscalar $^{56}$Ni,
and spin-unsaturated isovector $^{48}$Ca systems,
and is subsequently verified by systematic calculations of the SO splittings
in magic nuclei. In Sect.~\ref{sec5} we discuss an impact of these changes
on binding energies in magic nuclei. Conclusions of the paper are presented in
Sect.~\ref{sec6}.

\section{Spin-orbit and tensor energy densities, mean fields, and
interactions}
\label{sec2}

We begin by recalling the form of the EDF, which will be
used in the present study. In the notation defined in
Ref.~\cite{[Dob95e]} (see Ref.~\cite{[Per04]} for more details and
extensions), the EDF reads
   \begin{equation}\label{eq107}
   {\mathcal E} = \int {\rm d}^3{\mathbf r}\, {\mathcal H}({\mathbf r}),
   \end{equation}
where the local energy density ${\mathcal H}({\mathbf r})$ is
a sum of the kinetic energy, and of the
potential-energy isoscalar ($t=0$) and isovector ($t=1$) terms,
   \begin{equation}\label{eq108}
   {\mathcal H}({\mathbf r}) = \frac{\hbar^2}{2m}\tau_0
               + {\mathcal H}_0({\mathbf r})
               + {\mathcal H}_1({\mathbf r}) ,
   \end{equation}
with
   \begin{equation}\label{eq109a}
   {\mathcal H}_t({\mathbf r})  =  {\mathcal H}^{\text{even}}_t({\mathbf r})
                                +  {\mathcal H}^{\text{odd}}_t ({\mathbf r}) ,
   \end{equation}
and
\begin{eqnarray}
\label{hte}
\mathcal{H}_t^{\text{even}}
& = & C^{\rho}_t \rho^2_t + C^{\Delta \rho}_t
\rho_t\Delta\rho_t + \\ \nonumber
&\quad & C^{\tau}_t
\rho_t\tau_t + C^J_t {\mathbb J}^2_t +
C^{\nabla J}_t \rho_t {\mathbf \nabla}\cdot{\mathbf  J}_t,
\end{eqnarray}
\begin{eqnarray}
\label{hto}
\mathcal{H}_t^{\text{odd}} & = &C^{s}_t {\mathbf s}^2_t
+ C^{\Delta s}_t {\mathbf s}_t\cdot\Delta {\mathbf s}_t +  \\ \nonumber
&\quad &C^{T}_t{\mathbf s}_t \cdot {\mathbf T}_t +
C^j_t {\mathbf j^2_t} +
C^{\nabla J}_t {\mathbf s}_t \cdot ({\mathbf \nabla}\times {\mathbf j}_t).
\end{eqnarray}
For the time-even, $\rho_t$, $\tau_t$, and ${\mathbb J}_t$,  and
time-odd, ${\mathbf s}_t$, ${\mathbf T}_t$, and ${\mathbf j}_t$, local
densities we follow the convention introduced in Ref.~\cite{[Eng75]},
see also Refs.~\cite{[Ben03],[Per04]} and references cited therein.

In particular, the SO density
${\mathbf  J}$ is the vector part of the spin-current tensor density
${\mathbb J}$, i.e.,
\begin{equation}
\label{spin-current}
{\mathbb J}_{\mu \nu} = \tfrac{1}{3} J^{(0)} \delta_{\mu \nu}
+  \tfrac{1}{2} \varepsilon_{\mu \nu \eta}
{J}_\eta  +    J^{(2)}_{\mu \nu},
\end{equation}
with
\begin{equation}
\label{spin-current2}
{\mathbb J}^2
\equiv \sum_{\mu \nu} {\mathbb J}_{\mu \nu}^2 =
\tfrac{1}{3} (J^{(0)})^2  +  \tfrac{1}{2} {\mathbf J}^2
+ \sum_{\mu \nu  } (J^{(2)}_{\mu \nu})^2  .
\end{equation}

In the context of the present study, the time-even tensor and SO parts of the EDF,
\begin{eqnarray}
\label{htta}
\mathcal{H}_T    & = & C^J_0 {\mathbb J}^2_0 +
                       C^J_1 {\mathbb J}^2_1 , \\
\label{httb}
\mathcal{H}_{SO} & = & C^{\nabla J}_0 \rho_0 {\mathbf \nabla}\cdot{\mathbf  J}_0  +
                       C^{\nabla J}_1 \rho_1 {\mathbf \nabla}\cdot{\mathbf  J}_1 ,
\end{eqnarray}
are of particular interest.

In the spherical-symmetry limit, the scalar $J^{(0)}$
and symmetric-tensor $J^{(2)}_{\mu \nu}$ parts of the spin-current tensor vanish,
and thus
\begin{eqnarray}
\label{httas}
\mathcal{H}_T    & = & \tfrac{1}{2} C^J_0 J^2_0(r) +
                       \tfrac{1}{2} C^J_1 J^2_1(r) , \\
\label{httbs}
\mathcal{H}_{SO} & = & -C^{\nabla J}_0 J_0(r)\frac{d\rho_0}{dr}
                       -C^{\nabla J}_1 J_1(r)\frac{d\rho_1}{dr} ,
\end{eqnarray}
where the SO density has only the radial component,
${\mathbf J}_t=\frac{{\mathbf r}}{r}J_t(r)$.
Variation of the tensor and SO parts of the EDF over the radial SO densities
$J(r)$ gives the spherical isoscalar ($t=0$) and isovector ($t=1$) SO
MFs,
\begin{eqnarray}\label{sot} W_t^{SO} &
= & \frac{1}{2r}\left( C^J_t J_t(r) - C^{\nabla J}_t \frac{d\rho_t}{dr}\right)
{\mathbf L} \cdot {\mathbf S},
\end{eqnarray}
which can be easily translated into the neutron ($q=n$) and proton ($q=p$) SO
MFs,
\begin{eqnarray}\label{sotnp} W_q^{SO} &=&
    \frac{1}{2r}\bigg\{(C^J_0-C^J_1) J_0(r) + 2C^J_1 J_q(r)
\\ \nonumber
                  &&-(C^{\nabla J}_0-C^{\nabla J}_1) \frac{d\rho_0}{dr}
                    - 2C^{\nabla J}_1 \frac{d\rho_q}{dr}\bigg\}
{\mathbf L} \cdot {\mathbf S}.
\end{eqnarray}
Although below we perform calculations without assuming spherical and
time-reversal symmetries, here we do not repeat general expressions
for the SO mean-fields, which can be found in
Refs.~\cite{[Eng75],[Per04]}. We also note that, in principle, in
this general case, one could use different coupling constant
multiplying each of the three terms in Eq.~(\ref{spin-current2}). In
the present exploratory study, we do not implement this possible
extension of the EDF, and we use unique tensor coupling constants
$C^J_t$, as defined in Eq.~(\ref{hte}).

Identical potential-energy terms of the EDF, Eqs.~(\ref{hte}) and (\ref{hto}),
are obtained by averaging the
Skyrme effective interaction within the Skyrme-Hartree-Fock (SHF)
approximation \cite{[Ben03]}. By this procedure, the EDF coupling
constants $C_t$ can be expressed through the Skyrme-force parameters,
and one can use parameterizations existing in the literature. It is
clear that one can study tensor and SO effects entirely within the
EDF formalism, i.e., by considering the corresponding tensor and SO
parts of the EDF,  Eqs.~(\ref{htta}) and (\ref{httb}), and coupling constants
$C^J_t$ and $C^{\nabla J}_t$, respectively. However, in order to link this approach
to those based on the Skyrme interactions, we recall here expressions
based on averaging the zero-range tensor and SO forces \cite{[Flo75],[Sta77]}, see
also Refs.~\cite{[Per04],[dob06c],[Les07]} for recent
analyses. Namely, in the spherical-symmetry limit, one has
\begin{equation}\label{h_t}
{\mathcal H}_T = \tfrac{5}{8} \left[ t_{\text{e}} J_n(r) J_p(r) +
 t_{\text{o}} (J_0^2(r) - J_n(r) J_p(r) ) \right] ,
\end{equation}
\begin{equation}\label{h_so}
{\mathcal H}_{SO} = \frac{1}{4}\left[3W_0J_0(r)\frac{d\rho_0}{dr}
+W_1J_1(r)\frac{d\rho_1}{dr}\right],
\end{equation}
where in Eq.~(\ref{h_so}) two different coupling constants, $W_0$ and
$W_1$, were introduced following Ref.~\cite{[Rei95]}.

The corresponding SO MFs read
\begin{eqnarray}\label{so} W_q^{SO} &
= & \frac{1}{2r}\bigg\{\tfrac{5}{8} \left( (t_{\text{e}} + t_{\text{o}})
J_0(r) - (t_{\text{e}} - t_{\text{o}}) J_q(r) \right)  \\ \nonumber
  & & +  \frac{1}{4}\left( (3W_0-W_1) \frac{d\rho_0}{dr} - 2W_1
\frac{d\rho_q}{dr}\right) \bigg\}
{\mathbf L} \cdot {\mathbf S} ,
\end{eqnarray}
[note that in Ref.~\cite{[dob06c]}, the factor of $\tfrac{1}{2}$ was
missing at the $W_0$ term of Eq.~(4)]. By comparing Eqs.~(\ref{sotnp})
and (\ref{so}), one obtains the following relations between the
coupling constants:
\begin{eqnarray}\label{cto}
C_0^J & = & \tfrac{5}{16}( 3t_{\text{o}} + t_{\text{e}} ) ,\\
C_1^J & = & \tfrac{5}{16}(  t_{\text{o}} - t_{\text{e}} ) , \label{cte} \\
C^{\nabla J}_0 & = & -\tfrac{3}{4} W_0 ,\\
C^{\nabla J}_1 & = & -\tfrac{1}{4} W_1.
\end{eqnarray}
For further discussion of the Skyrme forces and their relation to
tensor components we refer the reader to an extensive and complete
recent discussion presented in Ref.~\cite{[Les07]}.

In this exploratory work, we base our considerations on the EDF
method and deliberately break the connection between the functional
(\ref{hte}), and the Skyrme central, tensor, and SO forces.
Nevertheless, in the time-even sector, our starting point is the
conventional Skyrme-force-inspired functional with coupling constants fixed
at the values characteristic for either
SkP~\cite{[Dob84]}, SLy4~\cite{[Cha97fw]}, or SkO~\cite{[Rei99fw]}
Skyrme parameterizations. However, poorly known coupling constants in
the time-odd sector (those which are not related to the time-even ones
through the local-gauge invariance \cite{[Dob95e]}) are fixed
independently of their Skyrme-force values. For this purpose, the
spin coupling constants $C^{s}_t$ are readjusted to reproduce
empirical values of the Landau parameters, according to the
prescription given in Refs.~\cite{[Ben02afw],[Zdu05xw]}, and
$C^{\Delta s}_t$ are set equal to zero. These variants of the standard
functionals are below denoted by SkP$_{L}$, SLy4$_{L}$, and SkO$_{L}$.

Strictly pragmatic reasons, like technical complexity and lack of
firm experimental benchmarks, made the majority of older Skyrme
parameterizations simply disregard the tensor terms, by setting
$C^J_t\equiv 0$. Recent experimental discoveries of new magic
shell-openings in neutron-reach light nuclei, e.g., around $N=32$
\cite{[For04],[Din05]}, and their subsequent interpretation in terms
of tensor interaction within the
shell-model~\cite{[Ots01],[Ots05],[Hon05]}, caused a revival of
interest in tensor terms within the MF approach
\cite{[Ots06],[dob06c],[Bro06],[Dob07b],[Bri07w],[Les07]}, which is
naturally tailored to study s.p.\ levels. Indeed, as shown in
Ref.~\cite{[dob06c]}, the tensor terms mark clear and unique
fingerprints in isotonic and isotopic evolution of s.p.\ levels and,
in particular, in the SO splittings.

In this paper, we perform systematic study of the SO splittings. The
goal is to resolve contributions to the SO MF (\ref{sot}) due
to the tensor and SO parts of the EDF, and to readjust the
corresponding coupling constants $C^J_t$ and $C^{\nabla J}_t$. It is
shown that this goal can be essentially achieved by studying the
$f_{7/2}-f_{5/2}$ SO splittings in three key nuclei: $^{40}$Ca,
$^{48}$Ca, and $^{56}$Ni. Before we present in Sec.~\ref{sec4} details
of the fitting procedure and values of the obtained coupling
constants, first we discuss effects of the core polarization and its
influence on the calculated and SO splittings.

\section{Core-polarization effects}
\label{sec3}

In spite of the fact that the s.p.\ levels belong to basic building
blocks of the MF methods, there is still a vivid debate concerning their
physical reality. The question whether they constitute only a set of
auxiliary quantities, or represent real physical entities that can be
inferred from experimental data, was never of any concern for methods
based on phenomenological one-body potentials. Indeed, these
potentials were bluntly fitted directly to reproduce the s.p.\ levels
deduced, in one way or another, from empirical information around
doubly-magic nuclei, see, e.g., Refs.~\cite{[Dud81],[Sch07a]}. In
turn, these potentials also appear to properly (satisfactorily)
reproduce the one-quasiparticle band-heads in open-shell nuclei, see,
e.g., Ref.~\cite{[Naz90]}. This success seems to legitimate the
physical reliability of the theoretical s.p.\ levels within the
microscopic-macroscopic approaches.

The debate concerns mostly the self-consistent MF approaches based on
the EDF methods or two-body effective interactions. The arguments
typically brought forward in this context underline the fact that the
self-consistent MFs are most often tailored to reproduce bulk
nuclear properties like masses, densities, radii, and certain
properties of nuclear matter. Consequently, the underlying interactions are
non-local with effective masses $m^\star/m \sim 0.8$~\cite{[Jeu76w]},
what in turn artificially lowers the density of s.p.\ levels around the
Fermi energy. It was pointed out by several
authors~\cite{[Ham76aw],[Ber80w],[Lit06w]} that restitution of physical
density of the s.p.\ levels around the Fermi energy can be achieved
only after inclusion of particle-vibration coupling, i.e., by going
beyond MF.

Within the effective theories, however, these arguments do not seem to be fully
convincing. First of all, an effective theory with properly chosen data
set used to fit its parameters should automatically {\it select\/}
physical value of the effective mass and readjust other parameters
(coupling constants) to this value. Examples of such {\it implicit\/}
scaling are well known for the SHF theory including: ({\it i\/})
explicit effective mass scaling of the coupling constants $C^s$ and $C^T$
through the fit to the spectroscopic Landau
parameters~\cite{[Ben02afw]}; ({\it ii\/}) direct dependence of the
isovector coupling constants $C_1^\tau$ and $C_1^\rho$ on the
isoscalar effective mass through the fit to the (observable) symmetry
energy strength~\cite{[Sat06w2]}; ({\it iii\/}) numerical indications
for the $m^\star$ scaling of the SO interaction inferred
from the $f_{7/2}$-$d_{3/2}$ splittings in $A\sim 44$
nuclei~\cite{[Zdu05xw]}. The differences between various
parameterizations of the Skyrme-force (or functional) parameters
rather clearly suggest that such an {\it implicit\/} $m^\star$
dependence  of functional coupling constants is a fact, which is,
however, neither well recognized nor understood so far.

Secondly, the use of effective interaction with parameters fitted at
the MF level is not well justified in beyond-MF approximation. Rather
unavoidable double counting results in such a case, and quantitative
estimates of level shifts resulting from such calculations need not
be very reliable. It is, therefore, quite difficult to accept the
viewpoint that unsatisfactory spectroscopic properties of, in
particular, modern Skyrme forces can be cured solely by going beyond
MF. On the contrary, the magnitude of discrepancies between the SHF
and experimental s.p.\ levels (see below) rather clearly suggest
that: ({\it i\/}) the data sets used to fit the force (or functional)
parameters are incomplete and ({\it ii\/}) the interaction/functional
should be extended. This is exactly the task undertaken in the present
exploratory work. We extend the conventional EDFs based on Skyrme
interactions by including tensor terms, and fit the corresponding
coupling constants to the SO splittings rather than to masses. The
preliminary goal is to improve spectroscope properties of
functionals, even at the expense of the quality in reproducing the
binding energies.

Empirical energy of a given s.p.\ neutron orbital can be deduced from
the difference between the ground-state energy of the doubly-magic core
with $N$ neutrons, $E_0(N)$, and energies, $E_p(N+1)$
or $E_h(N-1)$, of its odd neighboring isotopes having a
single-particle (p) or single-hole (h) occupying that orbital, i.e.,
   \begin{eqnarray}
   \epsilon_p(N) &=& E_p(N+1) - E_0(N),  \label{eq307a}  \\
   \epsilon_h(N) &=& E_0(N) - E_h(N-1),  \label{eq307b}
   \end{eqnarray}
see, for example, Refs.~\cite{[Isa02w],[Oro96w],[Sch07a]} and
referenced cited therein. Note that the total energies above are
negative numbers and {\em decrease} with increasing numbers of
particles, $E_h(N-1)>E_0(N)>E_p(N+1)$. Similarly, we can define
a measure of the neutron shell gap as the difference between
the lowest single-particle and highest single-hole energy,
   \begin{eqnarray}
   \Delta\epsilon_{\text{gap}}(N) &=& \epsilon_p(N) - \epsilon_h(N)   \\
             &=& E_p(N+1) + E_h(N-1) - 2E_0(N). \nonumber \label{eq308}
   \end{eqnarray}
Single-particle energies of proton orbitals and proton shell gaps are
defined in an analogous way.

Consistently with the empirical definitions, in the present paper, the
same procedure is used on the theoretical level \cite{[Rut99a]}. It means that we
determine the total energies of doubly-magic cores and their odd
neighbors by using the EDF method, and then we calculate the s.p.\
energies as the corresponding differences (\ref{eq307a}) or
(\ref{eq307b}). In this way, we avoid all the ambiguities related to
questions of what the s.p.\ energies really are and how they can be
extracted from data. Our methodology simply amounts to a specific way
of comparing measured and calculated masses of nuclei. In odd nuclei,
one has to ensure that particular s.p.\ orbitals are occupied by odd
particles, but still the calculated total energies correspond to
masses of their ground and low-lying excited states. Since here we
consider only doubly-magic nuclei and their odd neighbors, the
pairing correlations are neglected.

In our calculations, both time-even and time-odd polarizations
exerted by an odd particle/hole on the doubly-magic core are
evaluated self-consistently. In order to discuss these polarizations,
let us recall that the extreme s.p.\ model, in which ground-state
energies are sums of fixed s.p.\ energies of occupied orbitals, is
the model with no polarization of any kind included. In this model,
the differences of ground-state energies, Eqs.~(\ref{eq307a}) and
(\ref{eq307b}), are, of course, exactly equal to the model s.p.\
energies. Having this background model in mind, we can distinguish
three kinds of polarization effects:

\subsection{Mass polarization effect (time-even)}
\label{sec3a}

This effect can be understood as the self-consistent rearrangement of
all nucleons, which is induced by an added or subtracted odd
particle, while the shape is constrained to the spherical point and
the time-odd mean fields are neglected. One has to keep in mind, that
none of the s.p.\ orbitals of a spherical multiplet is spherically
symmetric, and therefore, when a particle in such a state is added to
a closed core, the resulting state of an odd nucleus cannot be
spherically symmetric, and thus cannot be
self-consistent. A self-consistent solution can only be
realized within the so-called filling approximation, whereupon one
assumes the occupation probabilities of all orbitals belonging to the
spherical multiplet of angular momentum $j$ and degeneracy $2j+1$, to
be equal to $v_j^2=1/(2j+1)$.

\begin{figure}
\includegraphics[width=0.4\textwidth, clip]{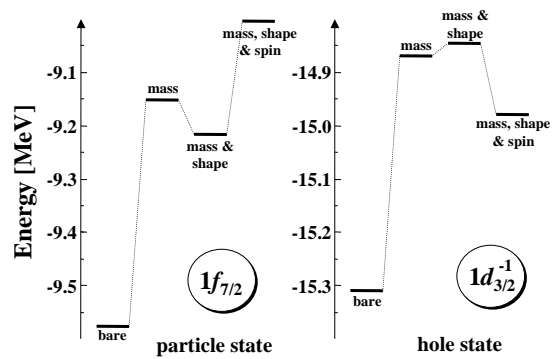}
\caption{Bare, mass polarized, mass and shape
polarized (time-even) and mass, shape and spin polarized s.p.\ energies
of the $\nu f_{7/2}$ particle
and the $\nu d_{3/2}^{-1}$ hole in $^{40}$Ca. The polarized s.p.\ levels
were deduced from binding energies in one particle/hole nuclei according to
the formulas~(\protect\ref{eq307a}) and (\protect\ref{eq307b}). Note that
attractive (repulsive) polarization corrections to the binding energies shift
the particle state down (up) and the hole state up (down), respectively.
}
\label{t0}
\end{figure}

The mass polarization effect is exactly zero when evaluated within
the first-order approximation. Indeed, as expressed by the Koopmans
theorem \cite{[Koo33]}, the linear term in variation of the total energy
with respect to adding or subtracting a particle is exactly equal to
the bare single-particle energy of the core. However, the obtained
results do not agree with Koopmans theorem. This is illustrated in
Fig.~\ref{t0}, where we compare bare and polarized s.p.\ energies of
the $\nu f_{7/2}$ and $\nu d_{3/2}$ orbitals in $^{40}$Ca. One can
see that for both orbitals there is a quite large and positive mass
polarization effect.

The reason for this disagreement lies in the fact that, because of the
center-of-mass correction, the standard EDF calculations are not really
variational with respect to adding or subtracting a particle.
Indeed, in these calculations, the total energy of an $A$-particle
system is corrected by the center-of-mass correction \cite{[Bei75],[Ben00d]},
\begin{equation}\label{eq710}
E_{\text{c.m.}} \simeq E^{\text{dir}}_{\text{c.m.}} =
  -\frac{1}{A}E_{\text{kin}} ,
\end{equation}
where $E_{\text{kin}}$ is the average kinetic energy of the system. The factor
of $\tfrac{1}{A}$ is not varied when constructing the standard
mean-field Hamiltonian, and therefore, the Koopmans theorem is
violated. Had this variation been included, it would have shifted
the mean-field potential and thus
all bare s.p.\ energies in given nucleus by a constant,
\begin{equation}\label{eq711}
\epsilon'_i = \epsilon_i
  +\frac{1}{A^2}E_{\text{kin}} .
\end{equation}
In $^{40}$Ca, this shift equals to 0.40\,MeV and almost exactly
corresponds to the entire mass polarization effect shown in
Fig.~\ref{t0}. The remaining mass polarization effect can be
attributed to higher-order polarization effects (beyond linear) and
to the non-selfconsistency of the filling approximation.

A few remarks about the shift in Eq.~(\ref{eq711}) are here in order.
First, the effect is independent on whether the two-body or one-body
center-of-mass correction \cite{[Ben00d]} is used, and on whether
this is done before or after variation. Only the detailed value of
the shift may depend on a particular implementation of the
center-of-mass correction. Second, the shift induces an awkward
result of the mean-field potential going asymptotically to a
positive constant and not to zero. Although this may seem to be a
trivial artifact, which does not influence the s.p.\ wave functions
and observables, it shows that the standard center-of-mass correction
should be regarded as an ill-defined theoretical construct. This fact
shows up as an acute problem in fission calculations \cite{[Ska06]}.
Third, whenever the bare s.p.\ energies are compared to empirical
data, this shift must by taken into account. Alternatively, as
advocated in the present study, one should compare directly the
calculated and measured mass differences. Finally, one should note
that the shift is irrelevant when differences of the s.p.\ energies
are considered, such as the SO splittings discussed in
Sec.~\ref{sec4}.

\subsection{Shape polarization effect (time-even)}

This effect is well known both in the MF and
particle-vibration-coupling approaches. Within a deformed MF theory
(with the time-odd mean fields neglected), it corresponds to a
simple fact that the s.p.\ energies (eigen-energies in a deformed
potential) depend on the deformation in specific way, which is
visualized by the standard Nilsson diagram \cite{[RS80]}. Indeed, in an axially
deformed potential, a spherical multiplet of angular momentum $j$
splits into $j+1/2$ orbitals according to moduli of the
angular-momentum projections $K=|m_j|$. Unless $j=1/2$, for prolate
and oblate deformations, orbitals with $K=1/2$ decrease and increase
in energy, respectively, while those with maximum $K=j$ behave in an
opposite way. Therefore, both for prolate and oblate deformations,
and for $j>1/2$, the lowest orbitals have the energies that are
lower than those at the spherical point. Hence, a $j>1/2$ particle
added to a doubly-magic core always polarizes the core in such a
way that the total energy decreases. On the other hand, the energy of
a $j=1/2, K=1/2$ orbital does not depend on deformation (in the first
order), and thus such an orbital does not exert any shape
polarization (in this order).

Exactly the same result is obtained in a particle-vibration-coupling
model, in which a $j>1/2$ particle can be coupled with either
0$^+$ or 2$^+$ state of the core, $[j\otimes0^+]_j$ or
$[j\otimes2^+]_j$, and the repulsion of these two configurations
decreases the energy of the ground state with respect to the unperturbed
spherical configuration $[j\otimes0^+]_j$. As before, for $j=1/2$, the
configuration $[j\otimes2^+]_j$ does not exist, and the ground state
is not lowered.

The above reasoning can be repeated for hole states, with the result
that the $j>1/2$ holes added to the doubly-magic core always
polarize the core in such a way that the total energy also
decreases. As a consequence, the shape polarization effect decreases
the s.p.\ energies of particle states (\ref{eq307a}) and increases
those of hole states (\ref{eq307b}), and thus {\em decreases} the shell gap
(\ref{eq308}). This effect is clearly illustrated in Fig.~\ref{t0}.

\subsection{Spin polarization effect (time-odd)}

When an odd particle or hole is added to the core, and the time-odd
fields are taken into account, it exerts polarization effects both in
the time-even (mass and shape) and time-odd (spin) channels. It means
that a non-zero average spin value of the odd particle induces a
time-odd component of the mean field, which influences average spin
values of all particles, leading to a self-consistent amplification of
the spin polarization.

It should be noted at this point that the spin polarization effect
dramatically depends on the assumed symmetries and choices made for
the occupied orbital, see discussion in the Appendix of
Ref.~\cite{[Olb06]}. Indeed, without the time-odd fields, in order to
occupy the odd particle or hole, one can use any linear combination
of states forming the Kramers-degenerate pair. The total energy is
independent of this choice, because the time-even density matrix
does not depend on it. This allows for making specific additional
assumptions about the conserved symmetries, e.g., in the standard
case, one assumes that the odd state is an eigenstate of the
signature (or simplex) symmetry with respect to the axis
perpendicular to the symmetry axis. However, in order to fully allow
for the spin polarization effects through time-odd fields, one has to
release all such restrictive symmetries and allow for alignment of
the spin of the particle along the symmetry axis. This requires
calculations with broken signature symmetry and only the
parity symmetry being conserved. Therefore, calculations of this kind
are more difficult than those performed within the standard cranking
model.

\subsection{Total polarization effect}

\begin{table*}
\begin{center}
\caption[A]{Neutron s.p.\ energies in $^{ 16}$O, $^{40}$Ca, and
$^{48}$Ca (in MeV). The
columns show: (a) bare unpolarized s.p.\ energies in doubly-magic
cores, (b) self-consistent s.p.\ energies obtained from binding
energies in one-particle/hole nuclei, Eqs.~(\protect\ref{eq307a}) and
(\protect\ref{eq307b}), with time-even mass and shape polarizations
included, (c) as in (b), but with time-odd spin polarizations
included in addition, and (d)
experimental data taken from Ref.~\cite{[Oro96w]}.
Time-even, (b)$-$(a), time-odd, (c)$-$(b), and total, (c)$-$(a) polarizations
are also shown, along with the differences between the
self-consistent and experimental spectra, (c)$-$(d). Positive s.p.\ energies
are shown only to indicate that particular orbitals are unbound in calculations;
their values are only very approximately related to positions of resonances.
All results have
been calculated using the Sly4$_{L}$ functional.}
\label{levels_sl4_1}
\begin{tabular}{c@{\mc}r@{\mc}r@{\mc}r@{\mc}r@{\mc}r@{\mc}r@{\mc}r@{\mc}r@{\mc}r}
\hline
\hline
          && \multicolumn{1}{c}{bare} & \multicolumn{1}{c}{T-even} & \multicolumn{1}{c}{T-even}    & \multicolumn{1}{c}{T-even}   & \multicolumn{1}{c}{T-odd}     & \multicolumn{1}{c}{total}     & \multicolumn{1}{c}{exp.}                    & \multicolumn{1}{c}{theory}    \\
          &&                          &                            & \multicolumn{1}{c}{pol.}      & \multicolumn{1}{c}{\& T-odd} & \multicolumn{1}{c}{pol.}      & \multicolumn{1}{c}{pol.}      & \multicolumn{1}{c}{\protect\cite{[Oro96w]}} & \multicolumn{1}{c}{$-$exp}    \\
          && \multicolumn{1}{c}{(a)}  & \multicolumn{1}{c}{(b)}    & \multicolumn{1}{c}{(b)$-$(a)} & \multicolumn{1}{c}{(c)}      & \multicolumn{1}{c}{(c)$-$(b)} & \multicolumn{1}{c}{(c)$-$(a)} & \multicolumn{1}{c}{(d)}                     & \multicolumn{1}{c}{(c)$-$(d)} \\
            \hline
$^{ 16}$O & $1\nu p_{ 3/2}$  & $-$20.57 & $-$19.61 &    0.96  & $-$20.29 & $-$0.68 &    0.28 & $-$21.84 &    1.55   \\
          & $1\nu p_{ 1/2}$  & $-$14.54 & $-$13.55 &    0.99  & $-$13.86 & $-$0.31 &    0.68 & $-$15.66 &    1.80   \\
            \cline{2-10}
          & $1\nu d_{ 5/2}$  &  $-$6.75 &  $-$5.83 &    0.92  &  $-$5.43 &    0.40 &    1.32 &  $-$4.22 & $-$1.21   \\
          & $2\nu s_{ 1/2}$  &  $-$3.78 &  $-$2.79 &    0.99  &  $-$2.30 &    0.49 &    1.48 &  $-$3.35 &    1.05   \\
          & $1\nu d_{ 3/2}$  &     0.39 &     1.19 &    0.80  &     1.37 &    0.18 &    0.98 &     1.50 & $-$0.13   \\
            \hline
$^{ 40}$Ca& $1\nu d_{ 5/2}$  & $-$22.01 & $-$21.55 &    0.46  & $-$21.87 & $-$0.32 &    0.14 & $-$22.39 &    0.52   \\
          & $2\nu s_{ 1/2}$  & $-$17.25 & $-$16.93 &    0.32  & $-$17.51 & $-$0.58 & $-$0.26 & $-$18.19 &    0.68   \\
          & $1\nu d_{ 3/2}$  & $-$15.31 & $-$14.84 &    0.47  & $-$14.98 & $-$0.14 &    0.33 & $-$15.64 &    0.66   \\
            \cline{2-10}
          & $1\nu f_{ 7/2}$  &  $-$9.58 &  $-$9.22 &    0.36  &  $-$9.00 &    0.22 &    0.58 &  $-$8.62 & $-$0.38   \\
          & $2\nu p_{ 3/2}$  &  $-$5.24 &  $-$4.85 &    0.39  &  $-$4.67 &    0.18 &    0.57 &  $-$6.76 &    2.09   \\
          & $2\nu p_{ 1/2}$  &  $-$3.06 &  $-$2.66 &    0.40  &  $-$2.55 &    0.11 &    0.51 &  $-$4.76 &    2.21   \\
          & $1\nu f_{ 5/2}$  &  $-$1.38 &  $-$1.10 &    0.28  &  $-$0.99 &    0.11 &    0.39 &  $-$3.38 &    2.39   \\
            \hline
$^{ 48}$Ca& $1\nu d_{ 5/2}$  & $-$22.60 & $-$22.02 &    0.58  & $-$22.21 & $-$0.19 &    0.39 & $-$17.31 & $-$4.90   \\
          & $2\nu s_{ 1/2}$  & $-$17.60 & $-$17.26 &    0.34  & $-$17.78 & $-$0.52 & $-$0.18 & $-$13.16 & $-$4.62   \\
          & $1\nu d_{ 3/2}$  & $-$16.55 & $-$15.97 &    0.58  & $-$16.02 & $-$0.05 &    0.53 & $-$12.01 & $-$4.01   \\
          & $1\nu f_{ 7/2}$  &  $-$9.79 &  $-$9.23 &    0.56  &  $-$9.34 & $-$0.11 &    0.45 &  $-$9.68 &    0.34   \\
            \cline{2-10}
          & $2\nu p_{ 3/2}$  &  $-$5.54 &  $-$5.25 &    0.29  &  $-$5.12 &    0.13 &    0.42 &  $-$5.25 &    0.13   \\
          & $2\nu p_{ 1/2}$  &  $-$3.54 &  $-$3.21 &    0.33  &  $-$3.11 &    0.10 &    0.43 &  $-$3.58 &    0.47   \\
          & $1\nu f_{ 5/2}$  &  $-$1.33 &  $-$1.25 &    0.08  &  $-$1.21 &    0.04 &    0.12 &  $-$1.67 &    0.46   \\
\hline
\hline
\end{tabular}
\end{center}
\end{table*}

\begin{table*}
\begin{center}
\caption[A]{Same as in Table \protect\ref{levels_sl4_1} but for
$^{90}$Zr, $^{132}$Sn, and $^{208}$Pb.}
\label{levels_sl4_2}
\begin{tabular}{c@{\mc}r@{\mc}r@{\mc}r@{\mc}r@{\mc}r@{\mc}r@{\mc}r@{\mc}r@{\mc}r}
\hline
\hline
          && \multicolumn{1}{c}{bare} & \multicolumn{1}{c}{T-even} & \multicolumn{1}{c}{T-even}    & \multicolumn{1}{c}{T-even}   & \multicolumn{1}{c}{T-odd}     & \multicolumn{1}{c}{total}     & \multicolumn{1}{c}{exp.}                    & \multicolumn{1}{c}{theory}    \\
          &&                          &                            & \multicolumn{1}{c}{pol.}      & \multicolumn{1}{c}{\& T-odd} & \multicolumn{1}{c}{pol.}      & \multicolumn{1}{c}{pol.}      & \multicolumn{1}{c}{\protect\cite{[Oro96w]}} & \multicolumn{1}{c}{$-$exp}    \\
          && \multicolumn{1}{c}{(a)}  & \multicolumn{1}{c}{(b)}    & \multicolumn{1}{c}{(b)$-$(a)} & \multicolumn{1}{c}{(c)}      & \multicolumn{1}{c}{(c)$-$(b)} & \multicolumn{1}{c}{(c)$-$(a)} & \multicolumn{1}{c}{(d)}                     & \multicolumn{1}{c}{(c)$-$(d)} \\
            \hline
$^{ 90}$Zr& $1\nu f_{ 7/2}$ & $-$23.16  & $-$22.83 &    0.33 & $-$22.94 & $-$0.11 &    0.22 & $-$14.76 & $-$8.18   \\
          & $1\nu f_{ 5/2}$ & $-$17.07  & $-$16.72 &    0.35 & $-$16.74 & $-$0.02 &    0.33 & $-$13.05 & $-$3.69   \\
          & $2\nu p_{ 3/2}$ & $-$17.52  & $-$17.30 &    0.22 & $-$17.44 & $-$0.14 &    0.08 & $-$12.74 & $-$4.70   \\
          & $2\nu p_{ 1/2}$ & $-$15.46  & $-$15.26 &    0.20 & $-$15.35 & $-$0.09 &    0.11 & $-$12.37 & $-$2.98   \\
          & $1\nu g_{ 9/2}$ & $-$12.08  & $-$11.75 &    0.33 & $-$11.81 & $-$0.06 &    0.27 & $-$11.69 & $-$0.12   \\
            \cline{2-10}
          & $2\nu d_{ 5/2}$ &  $-$6.73  &  $-$6.59 &    0.14 &  $-$6.52 &    0.07 &    0.21 &  $-$7.20 &    0.68   \\
          & $3\nu s_{ 1/2}$ &  $-$4.93  &  $-$4.70 &    0.23 &  $-$4.44 &    0.26 &    0.49 &  $-$5.78 &    1.34   \\
          & $2\nu d_{ 3/2}$ &  $-$3.99  &  $-$3.62 &    0.37 &  $-$3.59 &    0.03 &    0.40 &  $-$4.77 &    1.18   \\
          & $1\nu g_{ 7/2}$ &  $-$3.75  &  $-$3.75 &    0.00 &  $-$3.74 &    0.01 &    0.01 &  $-$4.62 &    0.88   \\
            \hline
$^{132}$Sn& $2\nu d_{ 5/2}$ & $-$11.72  & $-$11.48 &    0.24 & $-$11.56 & $-$0.08 &    0.16 &  $-$9.10 & $-$2.46   \\
          & $3\nu s_{ 1/2}$ &  $-$9.46  &  $-$9.28 &    0.18 &  $-$9.59 & $-$0.31 & $-$0.13 &  $-$7.55 & $-$2.04   \\
          & $1\nu h_{11/2}$ &  $-$7.66  &  $-$7.30 &    0.36 &  $-$7.33 & $-$0.03 &    0.33 &  $-$7.42 &    0.09   \\
          & $2\nu d_{ 3/2}$ &  $-$9.11  &  $-$8.91 &    0.20 &  $-$8.95 & $-$0.04 &    0.16 &  $-$7.17 & $-$1.78   \\
            \cline{2-10}
          & $2\nu f_{ 7/2}$ &  $-$2.01  &  $-$2.00 &    0.01 &  $-$1.95 &    0.05 &    0.06 &  $-$2.29 &    0.34   \\
          & $3\nu p_{ 3/2}$ &     0.17  &     0.26 &    0.09 &     0.31 &    0.05 &    0.14 &  $-$1.31 &    1.62   \\
          & $1\nu h_{ 9/2}$ &     0.95  &     0.79 & $-$0.16 &     0.77 & $-$0.02 & $-$0.18 &  $-$0.91 &    1.68   \\
          & $3\nu p_{ 1/2}$ &     0.97  &     1.08 &    0.11 &     1.12 &    0.04 &    0.15 &  $-$0.72 &    1.84   \\
          & $2\nu f_{ 5/2}$ &     0.82  &     0.84 &    0.02 &     0.88 &    0.04 &    0.06 &  $-$0.35 &    1.23   \\
            \hline
$^{208}$Pb& $2\nu f_{ 7/2}$ & $-$12.02  & $-$11.85 &    0.17 & $-$11.90 & $-$0.05 &    0.12 &  $-$9.96 & $-$1.94   \\
          & $1\nu i_{13/2}$ &  $-$9.52  &  $-$9.29 &    0.23 &  $-$9.30 & $-$0.01 &    0.22 &  $-$8.92 & $-$0.38   \\
          & $3\nu p_{ 3/2}$ &  $-$9.23  &  $-$9.09 &    0.14 &  $-$9.17 & $-$0.08 &    0.06 &  $-$8.12 & $-$1.05   \\
          & $2\nu f_{ 5/2}$ &  $-$9.03  &  $-$8.89 &    0.14 &  $-$8.91 & $-$0.02 &    0.12 &  $-$7.78 & $-$1.13   \\
          & $3\nu p_{ 1/2}$ &  $-$8.11  &  $-$8.01 &    0.10 &  $-$8.06 & $-$0.05 &    0.05 &  $-$7.72 & $-$0.34   \\
            \cline{2-10}
          & $2\nu g_{ 9/2}$ &  $-$3.19  &  $-$3.19 &    0.00 &  $-$3.16 &    0.03 &    0.03 &  $-$3.73 &    0.57   \\
          & $1\nu i_{11/2}$ &  $-$1.53  &  $-$1.65 & $-$0.12 &  $-$1.67 & $-$0.02 & $-$0.14 &  $-$3.11 &    1.44   \\
          & $3\nu d_{ 5/2}$ &  $-$0.50  &  $-$0.46 &    0.04 &  $-$0.43 &    0.03 &    0.07 &  $-$2.22 &    1.79   \\
          & $4\nu s_{ 1/2}$ &     0.56  &     0.65 &    0.09 &     0.80 &    0.15 &    0.24 &  $-$1.81 &    2.61   \\
          & $2\nu g_{ 7/2}$ &     0.08  &     0.10 &    0.02 &     0.11 &    0.01 &    0.03 &  $-$1.35 &    1.46   \\
          & $3\nu d_{ 3/2}$ &     0.69  &     0.76 &    0.07 &     0.78 &    0.02 &    0.09 &  $-$1.33 &    2.11   \\
\hline
\hline
\end{tabular}
\end{center}
\end{table*}

In Tables~\ref{levels_sl4_1} and \ref{levels_sl4_2} we list the
neutron s.p.\ energies calculated in six doubly-magic nuclei for the
SLy4$_{L}$ interaction. The bare s.p.\ energies (a) are compared to
those calculated from total energies, Eqs.~(\ref{eq307a}) and
(\ref{eq307b}), with the mass and shape (b) or mass, shape, and spin
(c) polarizations included.
In order to remove ambiguities associated with occupancy of the valence
particle (hole), binding energies of odd-$A$ nuclei were calculated by
blocking the lowest (highest) $K=j$ orbitals at oblate (prolate) shape
for particle
(hole) orbitals. The blocked orbitals were selected by performing cranking
calculation with angular-frequency vector parallel to the symmetry axis.
Such a cranking does not affect total energy or wave function, but
splits spherical multiplets into orbitals having good projections of
the angular momentum on the symmetry axis.
Calculations were performed by using the
code HFODD (v2.30a) \cite{[Dob05],[Dob05a],[Dob07c],[Dob07]} for the
spherical basis of $N_{\text{sh}}=14$ harmonic-oscillator shells.

As seen by comparing columns (b) and (a) of Tables~\ref{levels_sl4_1} and
\ref{levels_sl4_2}, the energy shifts caused by the time-even polarization
effects with respect to bare s.p.\ spectra are almost always positive, both
for particle and hole states. A few exceptions occur only for
large-$j$ unfavored ($j=\ell-1/2$) SO partners in heavy nuclei.
These shifts clearly decrease in magnitude
with increasing mass, from about 1\,MeV in $^{16}$O to below
0.25\,MeV in $^{208}$Pb. As discussed in Sec.~\ref{sec3a}, they are
mainly caused by the mass-polarization effects related to the center-of-mass
correction. Indeed, shifts of s.p.\ energies (\ref{eq711}), calculated
for the six doubly-magic nuclei of Tables~\ref{levels_sl4_1} and
\ref{levels_sl4_2}, are 0.87, 0.40, 0.36, 0.20, 0.14, and 0.09\,MeV,
respectively.

It is also
clearly visible that shifts of particle states are systematically
smaller than those of hole states, i.e., the time-even polarizations
tend to slightly {\em decrease} shell gaps.

The time-odd polarization effects systematically shift the hole
states down and particle states up in energy, i.e., they result in an
{\em increase} of shell gaps, cf.~also Fig.~\ref{t0}.
This result is at variance with that
obtained within the RMF approach \cite{[Rut99a]}, where the time-odd
fields corresponded to magnetic properties driven by the Lorentz
invariance, while here they are determined by experimental values of
the Landau parameters \cite{[Ben02afw],[Zdu05xw]}. We note here that
in recent derivations of the time-odd coupling constants within
the relativistic point-coupling model \cite{[Fin07]}, one obtains
values of the Landau parameters compatible with experimental values.
Shifts of s.p.\ energies due to the time-odd polarization effects
also decrease with mass, from about $-$0.7(+0.5)\,MeV in $^{16}$O to
below $-$0.1(0.15)\,MeV in $^{208}$Pb for hole (particle) states.

The total effect of combined time-even and time-odd polarizations
results in adding up the shifts for particle states and subtracting
those for hole states. In this way, the total shifts of particle and
hole states become mostly positive and (apart from light nuclei)
comparable in magnitude, with quite small net effects on shell gaps.
They also decrease with increasing mass, from up to 1.5\,MeV in
$^{16}$O to below 0.25\,MeV in $^{208}$Pb. Altogether, polarization
effects turn out to be significantly smaller than those obtained in
previous estimates. Although in quantitative analysis they cannot at
all be neglected, discrepancies with experimental data (last columns
in Tables~\ref{levels_sl4_1} and \ref{levels_sl4_2}) are still
markedly larger in magnitude. Therefore, bare s.p.\ energies can be
safely used, at least in all studies that do not achieve any better
overall agreement with data.

\section{Spin-orbit splittings}
\label{sec4}

Before proceeding to readjustments of coupling constants
so as to improve the agreement of the SO splittings with data,
we analyze the influence of time-even (mass and shape) and
time-odd (spin) polarization effects on the neutron SO splittings.
Based on results presented in the preceding Section, we calculate
the SO splittings as
   \begin{equation}\label{eq309}
   \Delta\epsilon_{\text{SO}}^{n{\ell}}  = \epsilon_{n{\ell}j_<}
                                         - \epsilon_{n{\ell}j_>}.
   \end{equation}

\begin{figure}
\includegraphics[width=0.48\textwidth, clip]{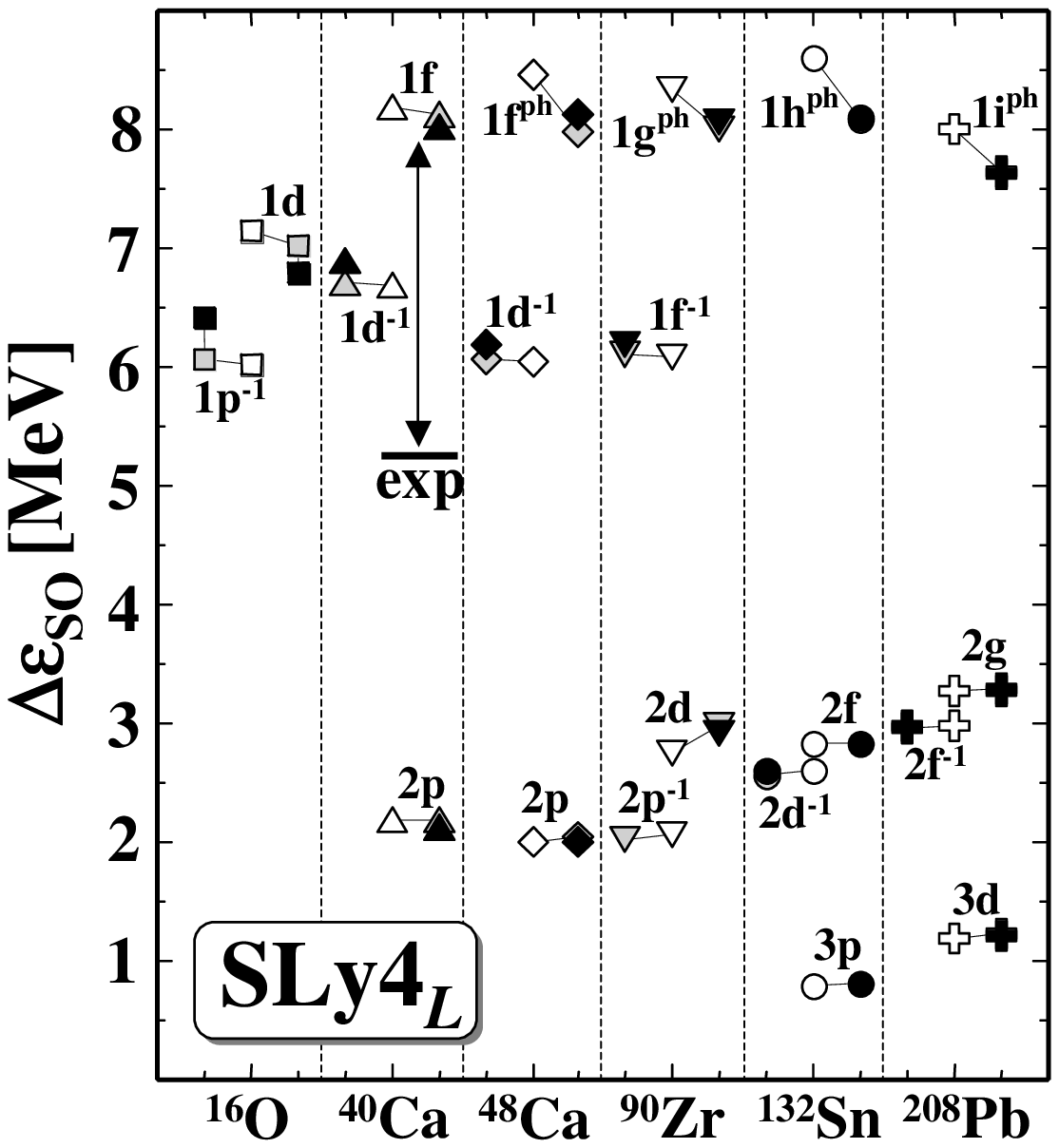}
\caption{Neutron SO splittings (\protect\ref{eq309}) calculated using
the SLy4$_{L}$ functional. White, gray, and black symbols mark bare,
mass and shape polarized (time-even), and mass, shape, and spin
polarized (time-even and time-odd) results, respectively. Results for
hole (particle and particle-hole) orbitals are shifted to the left
(right) with respect to the core (open symbols). A typical discrepancy
with experiment is shown by the arrow in $^{40}$Ca.
}\label{t1} \end{figure}

Figure~\ref{t1} shows the SO splittings calculated using SLy4$_{L}$ --- the
functional based on the original SLy4~\cite{[Cha97fw]} functional
with spin fields readjusted to reproduce empirical Landau parameters
according to the prescription given in
Refs.~\cite{[Ben02afw],[Zdu05xw]}. Plotted values correspond to
results presented in Tables~\ref{levels_sl4_1} and
\ref{levels_sl4_2}. The results are labeled according to the
following convention: open symbols mark results computed directly
from the s.p.\ spectra in doubly-magic nuclei (bare s.p.\ energies).
These bare values contain no polarization effect. Gray symbols label
the SO splittings involving polarization due to the time-even mass-
and shape-driving effects, i.e., those obtained with all time-odd
components in the functional set equal to zero. Black symbols
illustrate fully self-consistent results obtained for the complete
SLy4$_{L}$ functional. Gray and black symbols are shifted slightly to
the left-hand (right-hand) side with respect to the doubly-magic
core in order to indicate the hole (particle) character of the SO
partners. Mixed cases involving the particle-hole SO
partners are also shifted to the right.

The impact of polarization effects on the SO splittings is indeed
very small, particularly for the cases where both SO partners are of
particle or hole type. Indeed, for these cases, the effect only
exceptionally exceeds 200\,keV, reflecting a cancellation of
polarization effects exerted on the $j=\ell\pm 1/2$ partners. The
smallness of polarization effects hardly allows for any systematic
trends to be pinned down. Nevertheless, the self-consistent results
show a weak but relatively clear tendency to slightly enlarge
or diminish the splitting for hole or particle states, respectively.

The situation is clearer when the SO partners are of mixed particle
($j=\ell-1/2$) and hole ($j=\ell + 1/2$) character. In these cases,
the shape polarization tends to diminish the splitting quite
systematically by about 400--500\,keV. This behavior follows from
naive deformed Nilsson model picture where the highest-$K$ members of
the $j=\ell-1/2$ ($j=\ell+1/2$) multiplet slopes down (up) as a
function of the oblate (prolate) deformation parameter. As discussed
in the previous Section, the time-odd fields act in the opposite way,
tending to slightly enlarge the gap. The net polarization effect does
not seem to exceed about 300\,keV. In these cases, however, we deal
with large $\ell$ orbitals having also quite large SO splittings of
the order of $\sim$8\,MeV. Hence, the relative corrections due to
polarization effects do not exceed about 4\%, i.e., they are
relatively small -- much smaller than the effects of tensor terms
discussed below and the discrepancy with data, which in Fig.~\ref{t1}
is indicated for the neutron 1f SO splitting in $^{40}$Ca. These
results legitimate the direct use of bare s.p.\ spectra in magic
cores for further studies of the SO splittings, which considerably
facilitates the calculations.

As already mentioned, empirical s.p.\ energies are essentially
deduced from differences between binding energies of doubly-magic core
and their odd-$A$ neighbors. Different authors, however, use also one-particle
transfer data, apply phenomenological particle-vibration corrections
and/or treat slightly differently fragmented levels. Hence, published
compilations of the s.p.\ energies, and in turn the SO splittings, differ slightly
from one another depending on the assumed strategy. The typical
uncertainties in the empirical SO splittings can be inferred from
Table~\ref{sodata}, which summarizes the available data on the SO splittings
based on three recent s.p.\ level compilations published in
Refs.~\cite{[Isa02w],[Oro96w],[Sch07a]}.

\begin{table}
\begin{center}
\begin{tabular}{ccccc}
\hline
\hline
Nucleus &  orbitals &  Ref.~\protect\cite{[Isa02w]} &
            Ref.~\protect\cite{[Oro96w]} &  Ref.~\protect\cite{[Sch07a]}  \\
\hline
\hline
$^{16}$O & $\nu 1p_{3/2}^{-1}-\nu 1p_{1/2}^{-1}$ &    6.17  &  6.18  &  $--$\\
         & $\nu 1d_{5/2}- \nu 1d_{3/2}$          &    5.08  &  5.72  &  5.08\\
         & $\pi 1p_{3/2}^{-1}-\pi 1p_{1/2}^{-1}$ &    6.32  &  6.32  &  $--$\\
         & $\pi 1d_{5/2}- \pi 1d_{3/2}$          &    5.00  &  4.97  &  5.00\\
\hline
$^{40}$Ca& $\nu 2p_{3/2}- \nu 2p_{1/2}$          &    2.00  &  2.00  &  1.54\\
         & $\nu 1f_{7/2}- \nu 1f_{5/2}$          &    4.88  &  5.24  &  5.64\\
         & $\nu 1d_{5/2}^{-1}-\nu 1d_{3/2}^{-1}$ &    6.00  &  6.75  &  6.75\\
         & $\pi 2p_{3/2}- \pi 2p_{1/2}$          &    2.01  &  1.72  &  1.69\\
         & $\pi 1f_{7/2}- \pi 1f_{5/2}$          &    4.95  &  5.41  &  6.05\\
         & $\pi 1d_{5/2}^{-1}-\pi 1d_{3/2}^{-1}$ &    6.00  &  5.94  &  6.74\\
\hline
$^{48}$Ca& $\nu 2p_{3/2}- \nu 2p_{1/2}$          &  $--$    &  1.67  &  1.77\\
         & $\nu 1f_{7/2}^{-1}- \nu 1f_{5/2}$     &  $--$    &  8.01  &  8.80\\
         & $\nu 1d_{5/2}^{-1}-\nu 1d_{3/2}^{-1}$ &  $--$    &  5.30  &  3.08\\
         & $\pi 2p_{3/2}- \pi 2p_{1/2}$          &  $--$    &  2.14  &  1.77\\
         & $\pi 1f_{7/2}- \pi 1f_{5/2}$          &  $--$    &  4.92  &  $--$\\
         & $\pi 1d_{5/2}^{-1}-\pi 1d_{3/2}^{-1}$ &  $--$    &  5.01  &  5.29\\
\hline
$^{56}$Ni& $\nu 2p_{3/2}- \nu 2p_{1/2}$          &  $--$    &  1.88  &  1.12\\
         & $\nu 1f_{7/2}^{-1}- \nu 1f_{5/2}$     &  $--$    &  6.82  &  7.16\\
         & $\pi 2p_{3/2}- \pi 2p_{1/2}$          &  $--$    &  1.83  &  1.11\\
         & $\pi 1f_{7/2}^{-1}- \pi 1f_{5/2}$     &  $--$    &  7.01  &  7.50\\
\hline
$^{90}$Zr& $\nu 2d_{5/2}- \nu 2d_{3/2}$          &  $--$    &  2.43  &  $--$\\
         & $\nu 1g_{9/2}^{-1}- \nu 1g_{7/2}$     &  $--$    &  7.07  &  $--$\\
         & $\nu 2p_{3/2}^{-1}- \nu 2p_{1/2}^{-1}$&  $--$    &  0.37  &  $--$\\
         & $\nu 1f_{7/2}^{-1}- \nu 1f_{5/2}^{-1}$&  $--$    &  1.71  &  $--$\\
         & $\pi 2d_{5/2}- \pi 2d_{3/2}$          &  $--$    &  2.03  &  $--$\\
         & $\pi 1g_{9/2}- \pi 1g_{7/2}$          &  $--$    &  5.56  &  $--$\\
         & $\pi 2p_{3/2}^{-1}- \pi 2p_{1/2}^{-1}$&  $--$    &  1.50  &  $--$\\
         & $\pi 1f_{7/2}^{-1}- \pi 1f_{5/2}^{-1}$&  $--$    &  4.56  &  $--$\\
\hline
$^{100}$Sn& $\nu 2d_{5/2}- \nu 2d_{3/2}$          & 1.93  &  $--$ & 1.93\\
          & $\nu 1g_{9/2}^{-1}- \nu 1g_{7/2}$     & 7.00  &  $--$ & 7.00\\
          & $\pi 1g_{9/2}^{-1}- \pi 1g_{7/2}$     & 6.82  &  $--$ & 6.82\\
          & $\pi 2p_{3/2}^{-1}- \pi 2p_{3/2}^{-1}$& 2.85  &  $--$ & 2.85\\
\hline
$^{132}$Sn& $\nu 2f_{7/2}- \nu 2f_{5/2}$          & 2.00  &  1.94 & $--$\\
          & $\nu 3p_{3/2}- \nu 3p_{1/2}$          & 0.81  &  0.59 & 1.15\\
          & $\nu 1h_{11/2}^{-1}- \nu 1h_{9/2}$    & 6.53  &  6.51 & 6.68\\
          & $\nu 2d_{5/2}^{-1}- \nu 2d_{3/2}^{-1}$& 1.65  &  1.93 & 1.66\\
          & $\pi 2d_{5/2}- \pi 2d_{3/2}$          & 1.48  &  1.83 & 1.75\\
          & $\pi 1g_{9/2}^{-1}- \pi 1g_{7/2}$     & 6.08  &  5.33 & 6.08\\
\hline
$^{208}$Pb& $\nu 3d_{5/2}- \nu 3d_{3/2}$          & 0.97  &  0.89 & 0.97\\
          & $\nu 2g_{9/2}- \nu 2g_{7/2}$          & 2.50  &  2.38 & 2.50\\
          & $\nu 1i_{13/2}^{-1}- \nu 11_{11/2}$   & 5.84  &  5.81 & 6.08\\
          & $\nu 3p_{3/2}^{-1}- \nu 3p_{1/2}^{-1}$& 0.90  &  0.90 & 0.89\\
          & $\nu 2f_{7/2}^{-1}- \nu 2f_{5/2}^{-1}$& 2.13  &  2.18 & 1.87\\
          & $\pi 2f_{7/2}- \pi 2f_{5/2}$          & 1.93  &  2.02 & 1.93\\
          & $\pi 3p_{3/2}- \pi 3p_{1/2}$          & 0.85  &  0.45 & 0.52\\
          & $\pi 1h_{11/2}^{-1}- \pi 1h_{9/2}$    & 5.56  &  5.03 & 5.56\\
          & $\pi 2d_{5/2}^{-1}- \pi 2d_{3/2}^{-1}$& 1.68  &  1.62 & 1.46\\
\hline
\hline
\end{tabular}
\caption[A]{Empirical SO splittings. Compilation is based on the
empirical s.p.\ levels taken from Refs.~\cite{[Isa02w],[Oro96w],[Sch07a]}.
}
\label{sodata}
\end{center}
\end{table}

%-------------------------- F I G U R E 4 -----------------------------------
\begin{figure}
\includegraphics[width=0.4\textwidth, clip]{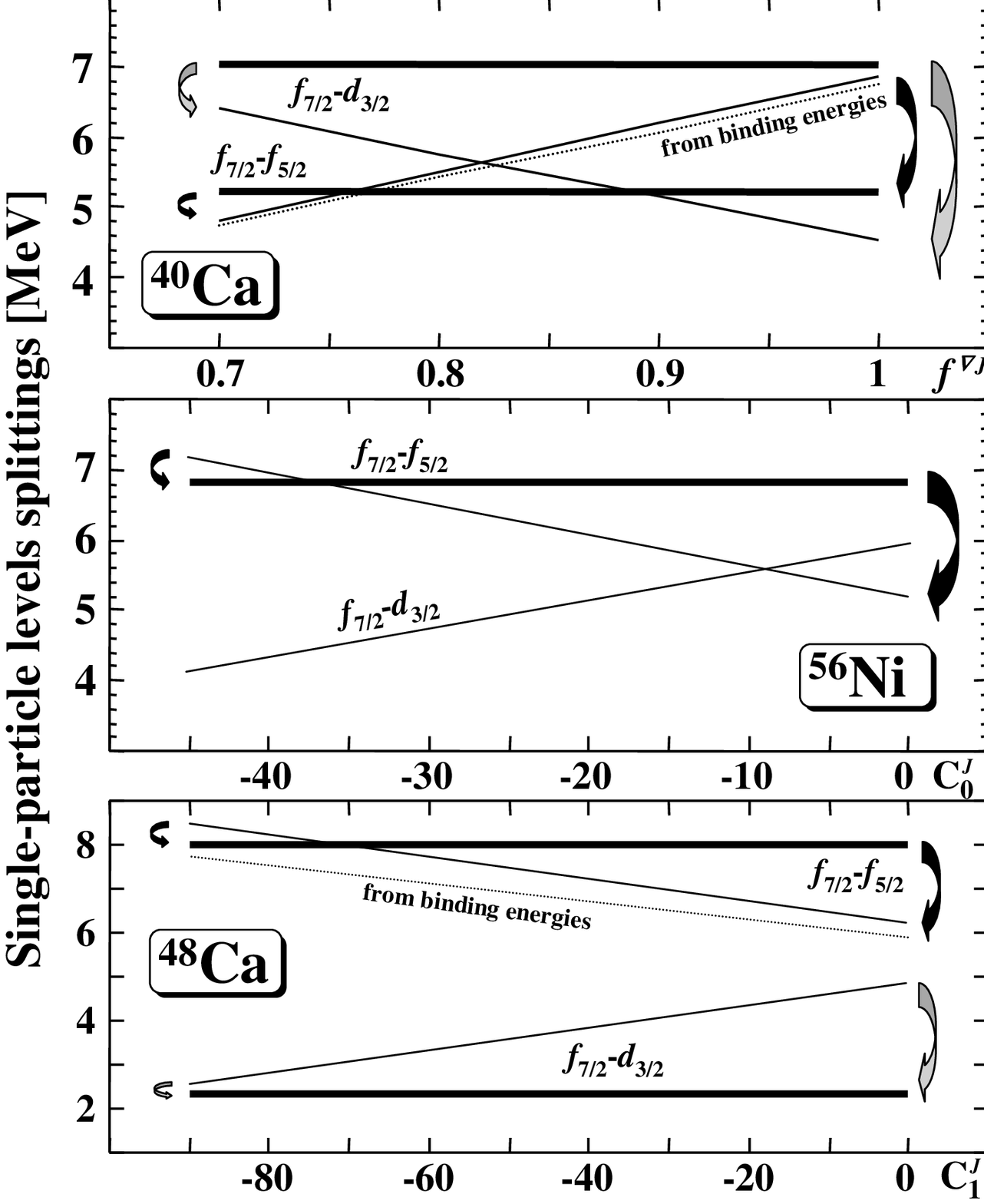}
\caption{Figure illustrates the three-step procedure used
to fit the isoscalar SO coupling constant $C^{\nabla J}_0$
in $^{40}$Ca (upper panel), the isoscalar tensor strength $C_0^J$
in $^{56}$Ni (middle panel), and the isovector tensor strength $C_1^J$
in $^{48}$Ca (lowest panel). These particular calculations have been done for
the SkP functional, but the pattern is common for all the
analyzed parameterizations including SLy4 and SkO. See text for
further details. }\label{t4} \end{figure}

Instead of large-scale fit to the data (see, e.g.,
Refs.~\cite{[Bri07w],[Les07]}), we propose a simple three-step method
to adjust three coupling constants $C_0^{\nabla J}$, $C_0^J$, and
$C_1^J$. The entire idea of this procedure is based on the
observation that the empirical $1f_{7/2}-1f_{5/2}$ SO splittings in
 $^{40}$Ca, $^{56}$Ni, and $^{48}$Ca form very
distinct pattern, which cannot be reproduced by using solely the
conventional SO interaction.

The readjustment is done in the following way. First, experimental
data in the spin-saturated (SS) nucleus $^{40}$Ca are used in order
to fit the isoscalar SO coupling constant $C^{\nabla J}_0$. One
should note that in this nucleus, the SO splitting depends {\em only}
on $C^{\nabla J}_0$, and not on $C^J_0$ (because of the spin
saturation), nor on $C^{\nabla J}_1$ (because of the isospin invariance at
$N=Z$), nor on $C^J_1$ (because of both reasons above). Therefore,
here one experimental number determines one particular coupling
constants.

Second, once $C^{\nabla J}_0$ is fixed, the spin-unsaturated (SUS)
$N=Z$ nucleus $^{56}$Ni is used to establish the isoscalar tensor
coupling constant $C_0^J$. Again here, because of the isospin
invariance, the SO splitting is independent of either of the two
isovector coupling constants, $C^{\nabla J}_1$ or $C^J_1$.  Finally,
in the third step, $^{48}$Ca is used to adjust the isovector tensor
coupling constant $C_1^J$. Such a procedure exemplifies the focus of
fit on the s.p.\ properties, as discussed in the Introduction.

It turns out that current experimental data, and in particular lack
of information in $^{48}$Ni, do not allow for adjusting the fourth
coupling constant, $C^{\nabla J}_1$. For this reason, in the present
study we fix it by keeping the ratio of $C^{\nabla J}_0/C^{\nabla
J}_1$ equal to that of the given standard Skyrme force. In the
process of fitting, all the remaining time-even coupling constants
$C_t$ are kept unchanged. Variants of the standard functionals
obtained in this way are below denoted by SkP$_{T}$, SLy4$_{T}$,
and SkO$_{T}$. When the time-odd channels, modified so as to reproduce
the Landau parameters, are active, we also use notation SkP$_{LT}$,
SLy4$_{LT}$, and SkO$_{LT}$.

For the SkP functional, the procedure is illustrated in Fig.~\ref{t4}.
We start with the isoscalar $N=Z$ nucleus $^{40}$Ca. The evolution of
the SO splittings in function of $f^{\nabla J}$, which is the factor
scaling the original SkP coupling constant $C_0^{\nabla J}$, is shown
in the upper panel of Fig.~\ref{t4}. As it is clearly seen from the
Figure, fair agreement with data requires about 20\% reduction in
the conventional SO interaction strength $C_0^{\nabla J}$. It should
be noted also that the reduction in the SO interaction
considerably improves the $1f_{7/2}-1d_{3/2}$ and $1f_{7/2}-2p_{3/2}$
splittings but slightly spoils the $2p_{3/2}-2p_{1/2}$ SO splitting.
Qualitatively, similar results were obtained for the SLy4 and SkO
interactions. Reasonable agreement to the data requires $\sim$20\%
reduction of the original $C_0^{\nabla J}$ in case of the SkO
interaction and quite drastic $\sim$35\% reduction
of the original $C_0^{\nabla J}$ in case of the SLy4 force.

Having fixed $C_0^{\nabla J}$ in $^{40}$Ca we move to the isoscalar
nucleus $^{56}$Ni.
This nucleus is spin-unsaturated and therefore is very sensitive
to the isoscalar $C_0^J$ tensor coupling constant.
The evolution of theoretical s.p.\ levels versus $C_0^J$ is illustrated
in the middle panel of Fig.~\ref{t4}.
As shown in the Figure, reasonable agreement between the empirical and
theoretical $1f_{7/2}-1f_{5/2}$ SO splitting is achieved
for  $C_0^J\sim -40$\,MeV\,fm$^5$, which by a factor of about five exceeds
the original SkP value for this coupling constant. It is striking
that a similar value of $C_0^J$ is obtained
in the analogical analysis performed for the SLy4 interaction.

Finally, the isovector tensor coupling constant $C_1^{\nabla J}$ is
established in $N$$\ne$$Z$ nucleus $^{48}$Ca.
The evolution of theoretical neutron s.p.\ levels versus
$C_1^J$ is illustrated in the lowest panel of Fig.~\ref{t4}.
As shown in the Figure, the value of  $C_1^{\nabla J}\sim -70$\,MeV\,fm$^5$
is needed to reach reasonable agreement for the $1\nu f_{7/2}-1\nu f_{5/2}$ SO
splitting in this case. For this value of the $C_1^J$ strength one
obtains also good agreement for the proton $1\pi f_{7/2}-1\pi f_{5/2}$ SO
splitting (see Figs.~\ref{t5} and \ref{t6} below), without any further
readjustment of the $C_1^{\nabla J}$ strength.
Again, very similar value for the $C_1^J$ strength is deduced
for the SLy4 force. Note also the improvement in the
$1f_{7/2}-1d_{3/2}$ splitting caused by the isovector
tensor interaction. Dotted lines show results obtained from the
mass differences, i.e., with all the polarization effects included.

During the fitting
procedure all the remaining functional
coupling constants were kept fixed at their Skyrme values. The ratio
of the isoscalar to the isovector coupling constant in the SO
interaction channel was locked to its standard Skyrme value of $C_0^{\nabla
J}/C_1^{\nabla J}=3$. Since no clear indication for relaxing this
condition is seen (see also Figs.~\ref{t5} and \ref{t6} below),
we have decided to investigate the isovector
degree of freedom in the SO interaction (see~\cite{[Rei95]})
by performing our three-step
fitting process also for the generalized Skyrme interaction
SkO~\cite{[Rei99fw]}, for which $C_0^{\nabla J}/C_1^{\nabla J}\sim -0.78$.

%\begingroup
%\squeezetable
\begin{table}
\begin{center}
\begin{tabular}{ccccc}
\hline
\hline
Skyrme &  $C_{0}^{\nabla J}$ &    $C_{0}^{\nabla J}/C_{1}^{\nabla J}$   &
$C_{0}^J$  & $C_{1}^J$ \\
 force &   [MeV\,fm$^5$]          &    & [MeV\,fm$^5$]    & [MeV\,fm$^5$]  \\
\hline
\hline
SkP$_{T}$    &   $-$60.0   &    3    &  $-$38.6      &  $-$61.7            \\
SLy4$_{T}$   &   $-$60.0   &    3    &  $-$45.0      &  $-$60.0            \\
\hline
SkO$_{T}$    &   $-$61.8   & $-$0.78 &  $-$33.1      &  $-$91.6            \\
\hline
\hline
\end{tabular}
\caption[A]{Spin-orbit $C^{\nabla J}$ and tensor isoscalar $C_{0}^J$  and
isovector $C_{1}^J$ functional coupling constants
adopted in this work and subsequently used in Figs.~\ref{t5}, \ref{t6}, and
\ref{t7}, where global calculations of the SO splittings
are presented.}
\label{fity}
\end{center}
\end{table}
%\endgroup

All the adopted functional coupling constants resulting from our
calculations are collected in Table~\ref{fity}.
Note, that the procedure leads to essentially identical
SO interaction strengths $C_0^{\nabla J}$
for all three forces irrespective of their intrinsic differences,
for example in effective masses. The tensor coupling constants in both
the SkP and the SLy4 functionals are also very similar. In the SkO case, one
observes rather clear enhancement in the isovector tensor coupling constant
which becomes more negative to, most likely, counterbalance the non-standard
positive strength in the isovector SO channel.

\begin{figure*}
\includegraphics[width=0.6\textwidth, clip]{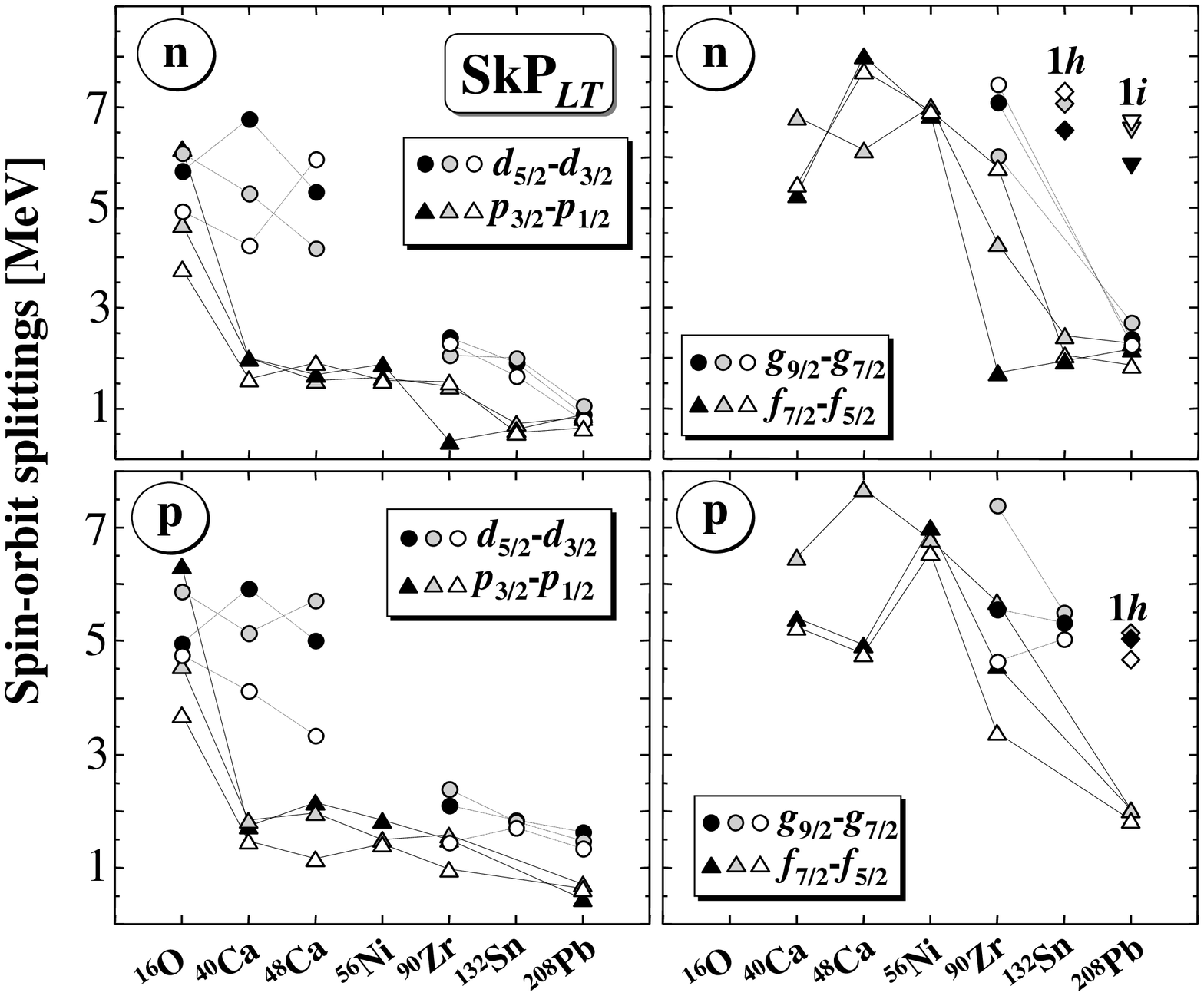}
\caption{Experimental~\protect{\cite{[Oro96w]}} (black symbols) and
theoretical SO splittings calculated using the original SkP functional
(gray symbols) and our modified SkP$_{LT}$ functional (white symbols)
with the SO and tensor coupling constants given in
Table~\protect\ref{fity}. Upper left and right panels show neutron
SO splittings for low-$\ell = 1,2$ ($p$ and $d$) and high-$\ell \geq 3$
orbitals, respectively. Analogical information but for
proton SO splittings is depicted in the lower panels.
}\label{t5}
\end{figure*}
%----------------------------------------------------------------------------
\begin{figure*}
\includegraphics[width=0.6\textwidth, clip]{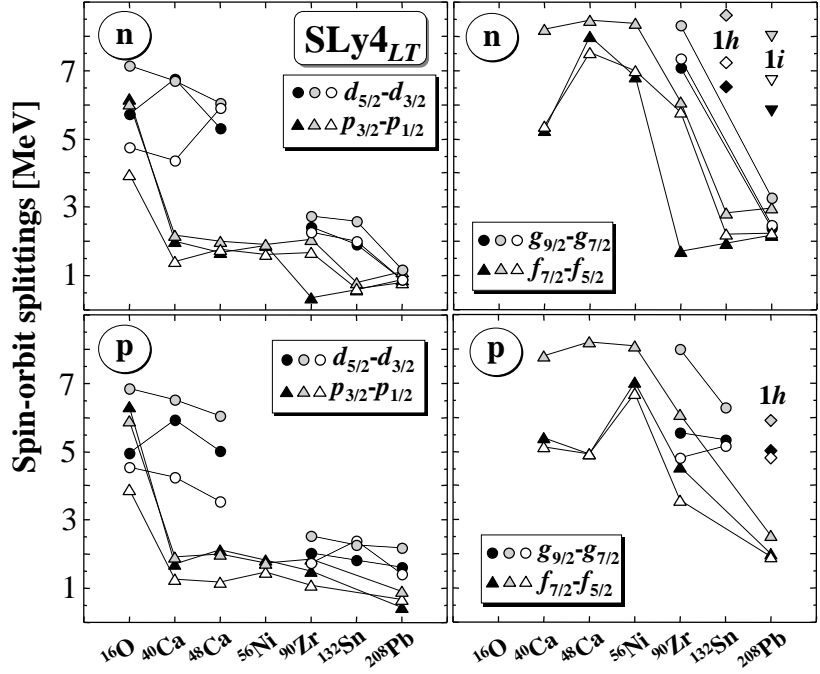}
\caption{Same as in Fig.~\ref{t5} but for the SLy4 functional.
}\label{t6}
\end{figure*}
%----------------------------------------------------------------------------
\begin{figure*}
\includegraphics[width=0.6\textwidth, clip]{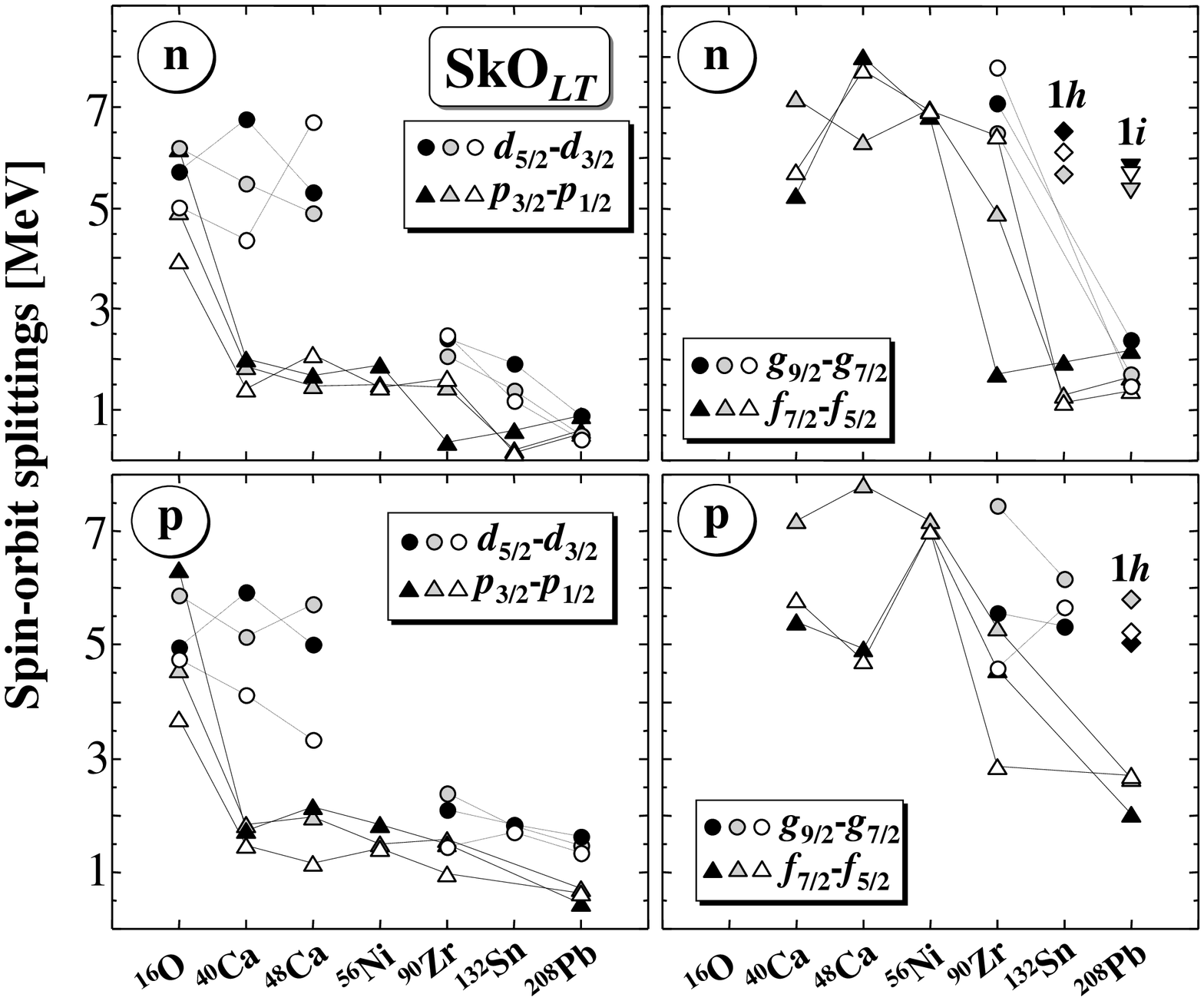}
\caption{Same as in Fig.~\ref{t5} but for the SkO functional.
}\label{t7}
\end{figure*}
%----------------------------------------------------------------------------

\begin{figure*}
\includegraphics[width=0.8\textwidth, clip]{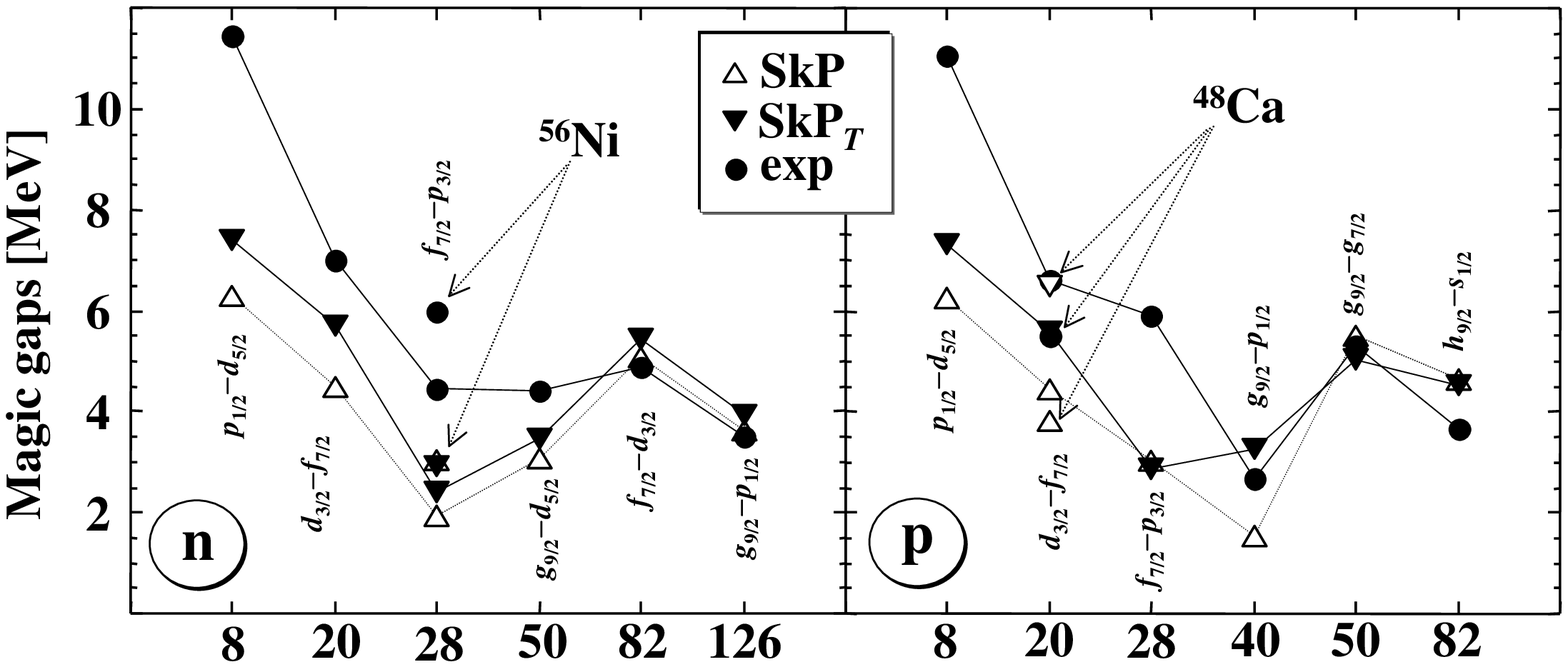}
\caption{Experimental~\protect{\cite{[Oro96w]}} (dots) and
theoretical values of magic gaps (\protect\ref{eq308}), calculated using the
original SkP functional (open triangles) and the SkP$_{T}$ functional
(full triangles) with the SO and the tensor coupling constants
from Table.~\protect\ref{fity}. The gaps were computed using the bare
unpolarized s.p.\ spectra. }
\label{t8}
\end{figure*}

The functionals were modified using only three specific pieces of data
on the neutron $1f_{7/2}-1f_{5/2}$ SO splittings. In order to verify the
reliability of the modifications,
we have performed systematic calculations of the
experimentally accessible SO splittings.
The results are depicted in Figs.~\ref{t5},
\ref{t6}, and \ref{t7}
for the SkP, SLy4, and SkO functionals,
respectively. Additionally, Fig.~\ref{t8} shows neutron and proton magic gaps (\ref{eq308})
calculated using the SkP functional.
In all these Figures, estimates taken from Ref.~\cite{[Oro96w]} are used as reference
empirical data.

This global set of the results can be summarized as
follows:
\begin{itemize}
\item
The $\ell=1$, $p_{3/2}-p_{1/2}$, SO splittings are slightly better reproduced
with original rather than modified functionals.
\item
The $1d$ SO splittings ($^{16}$O, $^{40,48}$Ca) are rather poorly reproduced
by both the original and modified functionals. These
splitting are also subject to relatively big empirical uncertainties
as shown in Table~\ref{sodata} and, therefore, need not be very conclusive.
In particular, the  $1d$ SO splittings in $N$=$Z$ $^{16}$O and $^{40}$Ca
nuclei deduced from Ref.~\cite{[Oro96w]} and depicted in the figures
show surprisingly large isospin dependence.
\item
The $2d$ and $3d$ splittings are quite well reproduced by both
the original and modified
functionals with slight preference for
the modified functional, in particular for the SLy4 interaction.
\item
All the $\ell\geq 3$ SO splittings are reproduced considerably better
by the modified functionals.
\item
Magic gaps are also better (although not fully satisfactorily) reproduced by
the modified functionals.
\end{itemize}

Without any doubt the SO splittings are better described by
the modified functionals. It should be stressed that the
improvements were reached using only three additional data points without
any further optimization. The tensor coupling constants
deduced in this work and collected in
table~\ref{fity} should be therefore considered as reference values.
Indeed, direct calculations show that variations in $C_t^J$ within $\pm$10\%
affect the calculated SO splittings only very weakly. The price paid for the
improvements concerns mostly the binding energies, which for the
nuclei $^{56}$Ni, $^{132}$Sn, and $^{208}$Pb become worse as compared
to the original values. This issue is addressed in the next section.

\section{Binding energies}
\label{sec5}

Parameters of the Skyrme functional have been fitted to reproduce
several physical quantities, with emphasis on the masses of magic nuclei.
Therefore it is not surprising that dramatic modifications of the
SO and tensor
terms of the functional, described in Sec.~\ref{sec3},
while improving the agreement between the calculated and measured
single-particle properties, can destroy the quality of the mass fit.
Hence, it is interesting to know whether this disagreement is significant
and whether it can be healed by refitting the remaining
parameters of the functional.

Table~\ref{tab:minim} shows differences between calculated and
experimental (Ref.~\cite{[Aud03]})
ground-state energies, $E_{\mathrm{calc}}-E_{\mathrm{exp}}$, (in MeV)
for a~set of spherical nuclei. Negative values mean that nuclei are overbound.
Results given in the second column,
denoted as SLy4, correspond to the standard SLy4~\cite{[Cha97fw]}
parametrization. The third column, denoted as SLy4$_{T}$,
illustrates significant deterioration of the quality of fit when
parameters $C^{\nabla J}_0$, $C_0^J$
and $C_1^J$ are modified (see Sec.~\ref{sec3}). Values presented in the last column,
SLy4$_{T\mathrm{min}}$, were obtained by minimizing the \emph{rms}
of relative discrepancies between the calculated and measured masses
Values of $C^{\nabla J}_0$, $C_0^J$ and $C_1^J$
were kept fixed at their SLy4$_{T}$ values, while
minimization was performed by varying the remaining parameters of the
functional $t_i$ and $x_i$. Note that for the standard SLy4
functional, the tensor coupling constants are set equal to zero
independently of the values of $t_i$ and $x_i$. For the minimization,
we have used the same methodology, namely, the influence of
parameters $t_i$ and $x_i$ on tensor coupling constants was
disregarded.

It should be emphasized that no attempt has been made to find the
global minimum --- the minimization was purely local, in the vicinity
of the standard SLy4 values of the parameters $t_i$ and $x_i$. One can see that even
this very limited procedure can lead to significant reduction of
discrepancies, down to quite reasonable values (with an exception
of $^{56}$Ni nucleus).

\begin{table}
\begin{center}
    \begin{tabular}{crrr}
        \hline\hline
        \rule[-0.6em]{0pt}{1.8em}Nucleus
                    &   SLy4   & SLy4$_{T}$
                                       &  SLy4$_{T\mathrm{min}}$\\
        \hline\hline
        \rule{0pt}{1.2em}$^{40}$Ca
                    & $-$2.197  &  $-$1.830  & $-$5.775     \\
         $^{48}$Ca  & $-$1.912  &     5.039  & $-$0.279     \\
         $^{56}$Ni  &    0.625  &    15.138  &    9.127     \\
         $^{90}$Zr  & $-$1.845  &     7.492  & $-$3.032     \\
        $^{132}$Sn  & $-$0.660  &    19.898  &    2.222     \\
        $^{208}$Pb  &    0.822  &    24.910  & $-$3.048     \\
        \hline
    \end{tabular}
    \caption{Differences between calculated and experimental
             ground-state energies, $E_{\mathrm{calc}}-E_{\mathrm{exp}}$, (in MeV)
             for a~set of spherical nuclei. Column denoted as SLy4$_{T\mathrm{min}}$
             shows results obtained after
             (local) minimization with respect to parameters
             $t_i, x_i,\, i = 0,1,2,3$. See text for details.
            }
    \label{tab:minim}
\end{center}
\end{table}

It is worth noting that the resulting modifications of the $t_i$ and
$x_i$ parameters turned out to be very small. Table~\ref{tab:params}
shows the values of the $t_i$ and $x_i$ parameters in the standard SLy4
parametrization (second column) and those obtained as the result of
the minimization procedure (third column). The last column shows
relative changes of parameters (in percent). As one can see, they are
at most of the order of one percent. Nevertheless, even such small
changes were sufficient to improve significantly the agreement between
calculated and experimental masses.

We stress again that the refitting procedure is used here only for
illustration purposes and the global fit to masses must probably
include extended functionals and improved methodology. For example,
the Wigner energy correction \cite{[Wig37]} was not included in the fit, as it was neither
included in the fit of the SLy4 parametrization. This correction
alone may change the balance of discrepancies obtained for the
$N=Z$ and $N\neq Z$ nuclei, and strongly impact the results. Systematic
studies of these effects will be performed in the near future.

\begin{table}
\begin{center}
    \begin{tabular}{crrr}
        \hline\hline
        \rule[-0.6em]{0pt}{1.8em}param.
               &     SLy4    &      SLy4$_{T\mathrm{min}}$
                                             &   change (\%)     \\
        \hline\hline
        \rule{0pt}{1.2em}$t_0$
               & $-$2488.913  & $-$2490.00300  &    0.04          \\
        $t_1$  &     486.818  &     486.78460  & $-$0.01          \\
        $t_2$  &  $-$546.395  &  $-$545.35849  & $-$0.19          \\
        $t_3$  &   13777.000  &   13767.77776  & $-$0.07          \\
        $x_0$  &       0.834  &       0.83257  & $-$0.17          \\
        $x_1$  &    $-$0.344  &    $-$0.34227  & $-$0.50          \\
        $x_2$  &    $-$1.000  &    $-$0.99798  & $-$0.20          \\
        $x_3$  &       1.354  &       1.36128  &    0.54          \\
        \hline
    \end{tabular}
    \caption{Skyrme force parameters $t_i, x_i,\, i = 0,1,2,3$ of the
             standard SLy4 parametrization (second column) compared with those
             obtained from the minimization procedure
             described in the text (third column). The last column
             shows relative change of parameters (in percent).
            }
    \label{tab:params}
\end{center}
\end{table}

\section{Conclusions}
\label{sec6}

Applicability of functional-based self-consistent mean-field or
energy-density-functional methods to nuclear structure
is hampered by their unsatisfactory s.p.\ properties.
This fact seems to be a mere consequence of strategies used to select
datasets that were applied in the process of adjusting
free parameters of these effective theories.
In spite of the fact that the s.p.\ energies are at the heart of
these methods, the datasets are heavily oriented towards
reproducing bulk nuclear properties in large-$N$ limit, with
only a marginal influence of the s.p.\ levels or level splittings
in finite nuclei.

In this work we suggest a necessity of shifting attention
from bulk to s.p.\ properties and to look for spectroscopic-quality
EDF, even at the expense of deteriorating its quality in reproducing
binding energies.  Such a strategy requires well-defined empirical input
related to the s.p.\ energies, to be used directly in the fitting process.
We argue that odd-even mass differences around magic nuclei
not only provide unambiguous direct information about nuclear
s.p.\ energies but are also well anchored within the spirit of the EDF
formalism. Indeed, the theorems due to Hochenberg and Kohn~\cite{[Hoh64w]} and
Levy~\cite{[Lev76]}, see
also Refs.~\cite{[Gir07]},  imply existence of universal EDF capable,
at least in principle, treating ground-states energies of nuclei
exactly. One can, therefore, argue that this implies essentially exact treatment
of at least the lowest s.p.\ levels forming ground states in one-particle
(one-hole) odd-$A$ nuclei with respect to even-even cores or, alternatively,
almost exact description of core-polarization phenomena caused by odd
single-particle (single-hole).

An
attempt to refit the EDF is
preceded by a
systematic analysis of the s.p.\ energies and self-consistent core-polarization
effects within the state-of-the-art Skyrme-force-inspired EDF. Three major sources
of core-polarization, including mass, shape and spin (time-odd) effects, are
identified and discussed. The analysis is performed for even-even doubly-magic cores and
the lowest s.p.\ states in odd-$A$
one-particle(hole) nuclei. The discussion is supplemented by analysis of the
s.p.\ SO splittings.

New strategy in fitting the EDF is applied to the SO and tensor
parts of the nuclear EDF. Instead of
large-scale fit to binding energies we propose simple and intuitive three-step
procedure that can be used to fit the isoscalar strength of the
SO interaction as well as the isoscalar and isovector strengths of the
tensor interaction. The entire idea is based on the observation that the
$f_{7/2} - f_{5/2}$ SO splittings in spin-saturated isoscalar $^{40}$Ca,
spin-unsaturated isoscalar $^{56}$Ni, and spin-unsaturated isovector $^{48}$Ca
form distinct pattern that can neither be understood  nor reproduced based
solely on the conventional SO interaction.
The procedure indicates a clear need for major reduction
(from $\sim$20\% till $\sim$35\% depending on the parameterization)
of the SO strength and for strong tensor fields.
It is verified that the suggested changes lead to systematic
improvements of the functional performance concerning such s.p.\ properties
like the SO splittings or magic gaps. It is also shown
that destructive impact of these changes on the binding energies can be
improved, to a large extent, by relatively small refinements of the remaining
coupling constants.

\vspace{0.3cm}

This work was supported in part by the Polish Ministry of
Science and by the Academy of Finland and University of
Jyv\"askyl\"a within the FIDIPRO programme.

%\bibliography{tensor,C:/Actual/LaTeX/Latex.all/jacwit23}
%\bibliography{tensor,jacwit23}
%\bibliographystyle{unsrt}

\end{document}